\documentstyle[aps,epsfig]{revtex}
\begin{document}

\title{Potential turning points in cluster radioactivity}

\author{D.N. Basu\thanks{E-mail:dnb@veccal.ernet.in}}
\address{Variable  Energy  Cyclotron  Centre,  1/AF Bidhan Nagar,
Kolkata 700 064, India}
\date{\today }
\maketitle
\begin{abstract}

      Effects of various nuclear interaction potentials on the decay lifetimes and the turning points of the WKB action integral has been studied. The microscopic nuclear potential obtained by folding in the density distribution functions of the two clusters with a realistic effective interaction has also been used to calculate the turning points. Half lives of $\alpha$ and $^{20} O$ cluster emissions from $^{228} Th$ have been calculated within the superasymmetric fission model using various phenomenological and the microscopic double folding potentials. Calculations of half lives with the microscopic double folding potentials are found to be in good agreement with the observed experimental data. Present calculations put the superasymmetric fission model on a firm theoretical basis.

\end{abstract}



      Experimental observation of emission of $^{14}C$ cluster from $^{222}Ra$ \cite{r1} has opened up a new vista in nuclear physics. Since then a lot of experimental and theoretical studies have been done in order to  understand the physics of cluster radioactivity. Lifetimes of the cluster radioactivities of radioactive nuclei have been predicted theoretically using various nuclear potentials within the superasymmetric fission model (SAFM) \cite{r2,r3,r4} and the preformed cluster model (PCM) \cite{r5}. In the SAFM the barrier penetrabilities are calculated assuming two asymmetric clusters while in the PCM the cluster is assumed to be formed before it penetrates the barrier and its preformation probability is also included in the calculations. Though the physics of the two approaches is apparently different but, in fact, they are very similar. Interpreting the cluster preformation probability within a fission model as the penetrability of the pre-scission part of the barrier, it was shown \cite{r6} that the PCM is equivalent to the fission model. But theoretical calculations showed that the PCM is better applicable for lighter clusters whereas the SAFM is more apt for all cluster decays \cite{r7}.

      The PCM have been tried with various nuclear potentials and compared with existing experimental data from time to time. It did achieve a reasonable success for the alpha radioactivity (using a cos-hyperbolic form for the nuclear interaction potential) \cite{r8}. However, it was not much succesful even for a very limited number of heavier cluster decays. The SAFM calculations using proximity type potentials or semiempirical heavy ion  potentials obtained by fitting the elastic scattering data or other phenomenological nuclear potentials for interaction between the fragments do not reproduce the observed cluster radioactivity lifetimes successfully. The SAFM using a parabolic potential approximation for the nuclear interaction potential, which is a rather unusual interaction potential between the daughter and emitted clusters, has been found to provide reasonable estimates for the lifetimes of cluster radioactivity \cite{r4}. In the present work microscopically calculated nuclear interaction potentials have been used in the SAFM approach for calculating the lifetimes of cluster radioactive decays. The microscopic nuclear potentential has been obtained by double folding the cluster density distributions with realistic effective interaction. This procedure of obtaining nuclear interaction potential is quite fundamental in nature. The results obtained using microscopic nuclear potentials have been compared with those obtained using various other nuclear interaction potentials. 

       The energetics allow spontaneous emission of cluster $(A_e, Z_e)$ from the parent nucleus $(A, Z)$, leaving a residual cluster $(A_d, Z_d)$, only if the released energy  

\begin{equation}
 Q = M - ( M_e + M_d)
\label{seqn1}
\end{equation}
\noindent
is a positive quantity, where $M$, $M_e$ and $M_d$ are the atomic masses of the parent, the emitted cluster and the daughter nuclei, respectively, expressed in the units of energy. Correctness of predictions for possible decay modes therefore rests on the accuracy of the ground state masses of nuclei.

     In the present work the total interaction energy $E(R)$ has been evaluated using microscopic nuclear potential $V(R)$ along with the Coulomb potential over the entire domain of interaction. The microscopic nuclear potentials have been obtained by double folding in the densities of the fragments with the finite range realistic M3Y effective interacion as

\begin{equation}
 V(R) = \int \int \rho_1(\vec{r_1}) \rho_2(\vec{r_2}) v[|\vec{r_2} - \vec{r_1} + \vec{R}|] d^3r_1 d^3r_2 
\label{seqn2}
\end{equation}
\noindent
The density distribution used for the clusters has been chosen to be of the spherically symmetric form given by

\begin{equation}
 \rho(r) = \rho_0 / [ 1 + exp( (r-c) / a ) ]
\label{seqn3}
\end{equation}                                                                                                                                           \noindent     
where                        
 
\begin{equation}
 c = r_\rho ( 1 - \pi^2 a^2 / 3 r_\rho^2 ), ~~     r_\rho = 1.13 A^{1/3}  ~~   and ~~    a = 0.54 ~ fm
\label{seqn4}
\end{equation}
\noindent
and the value of $\rho_0$ is fixed by equating the volume integral of the density distribution function to the mass number of the cluster. For $\alpha$ cluster, the density distribution has the form $\rho(r) = 0.4229 exp( - 0.7024 r^2)$. The finite range M3Y effective interaction $v(s)$ appearing in the eqn. (2) is given by \cite{r9} 

\begin{equation}
 v(s) = 7999. \exp( - 4s) / (4s) - 2134. \exp( - 2.5s) / (2.5s)
\label{seqn5}
\end{equation}   
\noindent

      In the SAFM the half life of the parent nucleus against the split into a cluster and a daughter is calculated using the WKB barrier penetration probability. The assault frequecy $\nu$ is obtained from the zero point vibration energy $E_v = (1/2)\hbar\omega = (1/2)h\nu$. The half life $T$ of the parent nucleus $(A, Z)$ against its split into a cluster $(A_e, Z_e)$ and a daughter $(A_d, Z_d)$  is given by

\begin{equation}
 T = [(h \ln2) / (2 E_v)] [1 + \exp(K)]
\label{seqn6}
\end{equation}
\noindent
where the action integral $K$ within the WKB approximation is given by

\begin{equation}
 K = (2/\hbar) \int_{R_a}^{R_b} {[2\mu (E(R) - E_v - Q)]}^{1/2} dR
\label{seqn7}
\end{equation}
\noindent
Here $\mu = mA_eA_d/A$  is the reduced mass, m is the nucleon mass, and $E(R)$ is the total interaction energy of the two fragments separated by the distance R between the centres, which is equal to the sum of nuclear interaction energy, Coulomb interaction energy and the centrifugal barrier.  The amount of energy released in the process is $Q$ and $R_a$ and $R_b$ are the two turning points of the WKB action integral determined from the equations

\begin{equation}
 E(R_a)  = Q + E_v = Q' = E(R_b)
\label{seqn8}
\end{equation} 
\noindent
 
      This microscopic nuclear potential energy  is then used to calculate the total interaction energy $E(R)$ for use inside the WKB action integral. The two turning points of the action integral have been obtained by solving eqns.(8) using microscopic double folding potential given by eqn.(2) along with the Coulomb potential. Then the WKB action integral between the two turning points has been evaluated numerically  for calculating the half lives of the cluster decays. The zero point vibration energies used in the present calculations are same as that described by eqns.(5)  in reference \cite{r10}. The shell effects for every cluster radioactivity are implicitly contained in the zero point vibration energy due to its proportionality with the Q value, which is maximum when the daughter nucleus has a magic number of neutrons and protons. 

      In the analytical superasymmetric fission model (ASAFM) \cite{r2} calculations, the entire interaction region is divided into two distinct zones. In the overlapping zone, where the distances of separation between the centres of the two fragments are below the touching radius  $R_t$, a parabolic form for the nuclear interaction potential has been used and distances beyond the touching radius $R_t$ have been treated as a nuclear force free zone and only Coulomb potential plus the centrifugal barrier for the separated fragments have been considered. This potential can be conveniently described by

\begin{eqnarray}
 E(R) =&&Q+[(E_i-Q)][(R-R_i)/(R_t-R_i)]^2~~~~~~~~~~~~~~~~~~~~for~~~~R \leq R_t \nonumber\\
        =&&Z_e Z_d e^2/ R+\hbar^2l(l+1)/(2\mu R^2)~~~~~~~~~~~~~~~~~~~~~~~~~~~for~~~~R\geq R_t 
\label{seqn9}
\end{eqnarray}   
\noindent
where $R_n=1.2249n^{1/3}$, $n=A~or~{A_e}~or~{A_d}$, $R_i=R_A-R_{A_e} $ and $R_t=R_{A_d}+R_{A_e} $ and $ E_i=E(R_t)$. Treating the region beyond the touching radius as a nuclear force free zone and approximating the nuclear interaction potential to a parabolic form in the overlapping region yield  analytical expression for the WKB action integral \cite{r2}.  Although the overall uncertainty of this analytical superasymmetric fission model (ASAFM) was found to be small, neither the division of the interaction region into two distinct domains is justifiable nor the use of parabolic nuclear potential has much physical basis.

      In this work, a relative comparison of calculations for the turning points and half lives for the $\alpha$ and $^{20} O$ cluster decays from $^{228} Th$ using various nuclear interaction potentials have been presented. In table-1, the results of calculations of the turning points of the WKB action integral and the SAFM calculations of the $\alpha$ decay half life have been presented which are obtained, respectively, using the microscopic double folding (DF) potential, the parabolic potential used by the ASAFM \cite{r2}, the proximity potential \cite{r11,r12}, Christensen and Winther semiempirical heavy ion potential \cite{r13} fitted to elastic scattering data 
 
\begin{equation}
 V(R)=-50[(R_{A_e}R_{A_d}/(R_{A_e}+R_{A_d})]exp[(R_{A_e}+R_{A_d}-R)/d] 
\label{seqn10}
\end{equation}
\noindent
where~~~~~~~$d=0.63 fm$,~~~~$R_n=(1.233n^{1/3}-0.978n^{-1/3})fm$,~~with~$n={A_e}or{A_d}$ ~~~~~~~~~~~~~~~~~~~~~~~~~~~~~~~~~~~~~~~~~~~~~~~~~~~~~~~~~~~~~~~~~~~~~~~~~~~~~~~~~~~and the Igo potential \cite{r14,r15} 

\begin{equation}
 V(R)=-1100exp[-(R-1.17A_d^{1/3})/0.574] 
\label{seqn11}
\end{equation}
\noindent
used for the $\alpha$ decay.
\begin{table}
\caption{Comparison of results for potential turning points and $\alpha$ decay half life obtained using different potentials.}
\begin{tabular}{cccc}
Nuclear Potential&First turning point&Second turning point&Theoretical half Life \\ 
Name & (fm) & (fm)  & $log_{10}T(s)$  \\ \hline
DF & 6.71 & 41.47 &8.08          \\
Parabolic &  6.20&41.47& 7.85         \\
Proximity    &  8.90   & 41.47 &4.81       \\
Christensen and Winther &9.33   &41.47&   4.20        \\
Igo  & 17.27 &41.47& -6.74        \\ 
ASAFM(1986) &   &    &  8.27         \\  
ASAFM(1991) &   &    &  7.96        \\ \hline
$Q=5.53 MeV$&  &  &Expt.$logT(s)=7.781\pm0.001$    \\
$E_v(old)=0.5254MeV$&  &  &$E_v(new)=0.5779MeV$    \\ 
\end{tabular} 
\end{table}
In table-2, the results of calculations of the turning points of the WKB action integral and the SAFM calculations of the $^{20} O$ decay half life have been presented which are obtained, respectively, using the microscopic double folding (DF), the parabolic, the proximity, Christensen and Winther and Igo potentials.

\begin{table}
\caption{Comparison of results for potential turning points and $^{20} O$ decay half life obtained using different potentials.}
\begin{tabular}{cccc}
Nuclear Potential&First turning point&Second turning point&Theoretical half Life \\ 
Name & (fm) & (fm)  & $log_{10}T(s)$  \\ \hline
DF & 7.69 & 19.87 &21.05          \\
Parabolic &  5.79&19.87& 21.75         \\
Proximity    &  9.80   & 19.87 &8.86       \\
Christensen and Winther &10.75   &19.87&   4.17        \\
Igo  & 17.01 &19.67& -18.81        \\ 
ASAFM(1986) &   &    &  22.44         \\  
ASAFM(1991) &   &    &  21.95        \\ \hline
$Q=44.72 MeV$&  &  &Expt.$logT(s)=20.86\pm0.30$    \\
$E_v(old)=2.5072MeV$&  &  &$E_v(new)=2.7070MeV$    \\ 
\end{tabular} 
\end{table}
As obvious from tables-1,2 the second turning point is primarily due to the Coulomb interaction and are identical. A slight difference for the Igo potential shows the residual effect of this large nuclear potential. Though the first turning point using the parabolic nuclear potential of ASAFM and the microscopic DF potential are very different, the results for the decay half life are quite close and reasonably close to the experimental result shown in the table. A normalisation factor of 0.9 for the microscopic nuclear potential has been used. The results obtained using other nuclear potentials are quite bad. The old and the new zero-point vibration energies given by eqn.[11] of reference \cite{r3} and eqns.(5)  of reference \cite{r10} respectively, have been displayed in the last lines of the tables-1,2. The comparison of the ASAFM(1991) results for the logarithmic half life obtained evaluating the analytical expressions \cite{r2} and those obtained using the parabolic potential used in reference \cite{r2} inside the WKB action integral and then evaluating the action integral after calculating its turning points, shows that the results differ by one percent approximately. In both the cases zero-point vibration energies given by eqns.(5)  of reference \cite{r10} have been used. It is important to mention that the only difference between  ASAFM(1986) and ASAFM(1991) calculations is that they use zero-point vibration energies given by eqn.[11] of reference \cite{r3} and eqns.(5)  of reference \cite{r10} respectively.

\begin{figure}[h]
\eject\centerline{\epsfig{file=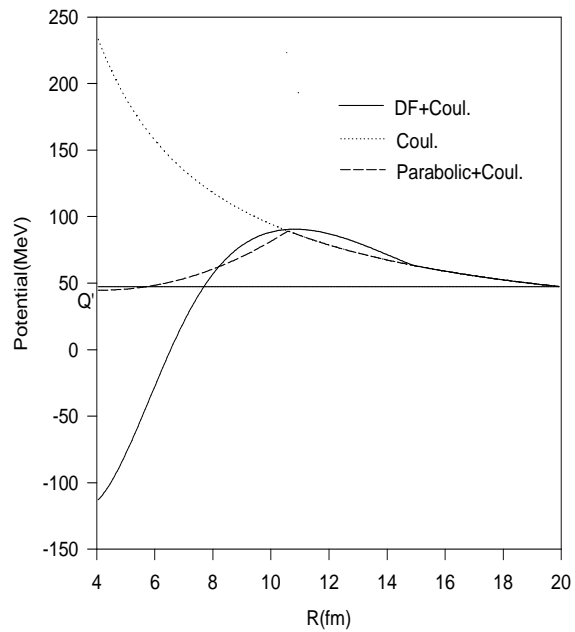,height=15cm,width=10cm}}
\caption
{ The nuclear DF potential with  a normalisation factor of 0.9 plus the Coulomb potential has been plotted as a function of radial separation between the fragments $^{20} O$ and $^{208} Pb$. The plot of the parabolic nuclear potential along with the Coulomb potential has also been shown in the figure. The Coulomb potential alone has also been shown.}
\label{fig1}
\end{figure}

      In Fig.~\ref{fig1} the nuclear DF potential with  a normalisation factor of 0.9 plus the Coulomb potential has been plotted as a function of radial separation between the fragments $^{20} O$ and $^{208} Pb$ . The plot of ASAFM parabolic nuclear potential along with the Coulomb potential which is given by eqn.(9) has also been shown in the figure for zero angular momentum between the fragments. The Coulomb potential alone has also been plotted. Obviously, beyond $R_t$ the ASAFM potential matches with the Coulomb potential because it was so designed. It is very interesting to observe that although the first turning point and the potentials between the turning points are so different for the parabolic potential of  ASAFM and the microscopic DF potential, the results for the half life are not much different. In table-3 and table-4, the variation with the normalisation factor used for the DF potential with the potential turning points and half lives of the $\alpha$ and $^{20} O$ cluster decays from $^{228} Th$ have been shown respectively. 

\begin{table}
\caption{Potential turning points and $\alpha$ decay half life obtained using different normalisation for DF potential.}
\begin{tabular}{cccc}
Normalisation&First turning point&Second turning point&Theoretical half Life \\ 
(no.) & (fm) & (fm)  & $log_{10}T(s)$ \\ \hline
1.0 & 7.05 & 41.47 &7.69          \\
0.9 &  6.71&41.47& 8.08     \\
0.8    &  6.37   & 41.47 &8.55       \\
0.7&5.98   &41.47& 9.14        \\
0.6  & 5.55 &41.47& 9.85        \\ \hline
LDM(1993)&    &    & 8.11  \\
$E_v(new)= 0.5779MeV$&  &  &Expt. $logT(s)=7.781\pm0.001 $    \\ 
\end{tabular} 
\end{table}

\begin{table}
\caption{Potential turning points and $^{20} O$ decay half life obtained using different normalisation for DF potential.}
\begin{tabular}{cccc}
Normalisation&First turning point&Second turning point&Theoretical half Life \\ 
(no.) & (fm) & (fm)  & $log_{10}T(s)$  \\ \hline
1.0 & 7.94 & 19.87 &20.22          \\
0.9 &  7.69&19.87& 21.05     \\
0.8    &  7.39   & 19.87 &22.33       \\
0.7&7.04   &19.87& 23.91        \\
0.6  & 6.49 &19.87& 26.44        \\ \hline
LDM(1993)&    &    & 21.69  \\
$E_v(new)= 2.7070MeV$&  &  &Expt. $logT(s)=20.86\pm0.30 $    \\ 
\end{tabular} 
\end{table}

      The result obtained by the liquid drop model (LDM) calculation \cite{r16} for the logarithmic decay half life for $\alpha$ is 8.11. The results obtained using the microscopic (DF) potential with normalisation 1.0 is 7.69 and with normalisation 0.9 is 8.08 and both the results are better than the LDM estimate. The result obtained by the liquid drop model (LDM) calculation \cite{r16} for the logarithmic decay half life for $^{20} O$  is 21.69 and the results obtained using the microscopic (DF) potential with normalisation 1.0 or 0.9 are 20.22 and 21.05 respectively and both the results are better than the LDM estimate. The normalisation of 0.9 has been found to provide optimum fit to the experimental results for the half lives of $\alpha$ decays and various heavy cluster decays \cite{r17}.  

\begin{table}
\caption{Comparison between Measured and Calculated $\alpha$-decay Half-Lives using the M3Y nucleon-nucleon effective interactions for the DF nuclear potential calculations. Calculations using normalisation constant C=0.9 for the DF nuclear interaction potentials have been presented. Corresponding results using the Viola-Seaborg relationship using Sobiczewski constants (VSS)  have also been presented. }
\begin{tabular}{cccccccccccc}
Parent &Parent &Released&VSS&DF&Expt.&Parent &Parent &Released &VSS&DF&Expt.      \\ 
           & &  energy &      &    &       &          &  & energy   &      &    &             \\
$Z$&$A$&$Q(MeV)$&$log_{10}T(s)$&$log_{10}T(s)$&$log_{10}T(s)$&$Z$&$A$&$Q(MeV)$&$log_{10}T(s)$&$log_{10}T(s)$&$log_{10}T(s)$ \\ \hline
       &       &         &  Expt.&   data&  from &reference& \cite{r16}.& & & & \\ 
   87&  221&   6.47&   2.76&   2.37&   2.50&   88&  221&   6.89&   1.82&   1.31&   1.45\\
   88&  222&   6.68&   1.58&   1.73&   1.58&   88&  223&   5.98&   5.68&   5.21&   6.00\\
   88&  224&   5.79&   5.53&   5.70&   5.50&   89&  225&   5.94&   6.05&   5.72&   5.94\\
   88&  226&   4.87&  10.70&  10.94&  10.70&   90&  228&   5.53&   7.86&   8.08&   7.78\\
   91&  231&   5.15&  11.34&  11.08&  12.01&   90&  230&   4.78&  12.40&  12.65&  12.38\\
   92&  232&   5.42&   9.49&   9.73&   9.34&   92&  233&   4.92&  13.63&  13.32&  12.70\\
   92&  234&   4.86&  12.97&  13.24&  12.89&   94&  236&   5.87&   8.04&   8.27&   7.95\\
   93&  237&   4.96&  13.62&  13.38&  13.83&   94&  238&   5.60&   9.49&   9.70&   9.44\\
   95&  241&   5.64&  10.54&  10.25&  10.14&   96&  242&   6.22&   7.22&   7.42&   7.15\\
       &      &          &Expt.&   data&  from&reference& \cite{r18}.& & & &\\ 
   90&  226&   6.46&   3.38&   3.55&   3.39&   90&  232&   4.08&  17.71&  18.04&  17.76\\
   92&  230&   6.00&   6.42&   6.65&   6.43&   92&  235&   4.69&  15.21&  14.90*&  16.57\\
   92&  236&   4.58&  14.94&  15.20&  14.99&   94&  240&   5.26&  11.47&  11.69&  11.45\\
    &      &              & Expt.&   data&  from&reference& \cite{r4}.& & & & \\ 
   54&  112&   3.33&   -.32&   1.95&   2.90&   70&  158&   4.18&   4.66&   5.75*&   3.90\\
   72&  160&   4.91&   1.68&   2.67&   2.70&   74&  164&   5.28&    .93&   1.85&   2.30\\
   80&  178&   6.58&  -1.58&   -.89&   -.20&   85&  215 &  8.18&  -3.92&  -4.36&  -4.00\\
   86&  215&   8.85&  -5.06&  -5.63&  -5.60&   86&  216&   8.20&  -4.37&  -4.24&  -4.30\\
   86&  217&   7.89&  -2.39&  -2.98&  -3.20&   86&  218&   7.27&  -1.46&  -1.33&  -1.40\\
   86&  219&   6.95&    .74&    .18&    .60&   86&  220 &  6.41&   1.78&   1.91&   1.70\\
   86&  222&   5.60&   5.49&   5.64&   5.50&   87&  216&   9.18&  -5.49&  -5.74&  -6.10\\
   87&  217&   8.47&  -3.98&  -4.40&  -4.60&   87&  218&   8.02&  -2.35&  -2.58&  -3.10\\
   87&  219&   7.46&   -.92&  -1.34&  -1.60&   87&  220&  6.81&   1.75&   1.57&   1.60\\
   88&  217&   9.16&  -5.14&  -5.68&  -5.80&   88&  218&   8.55&  -4.61&  -4.45&  -4.80\\
   88&  219&   8.13&  -2.34&  -2.89*&  -1.80&   88&  220&   7.60&  -1.75&  -1.60&  -1.60\\
   89&  217&   9.84&  -6.71&  -7.05&  -6.90&   89&  218&   9.38&  -5.29&  -5.50*&  -6.50\\
   89&  219&   8.83&  -4.23&  -4.62&  -5.10&   89&  220&   8.35&  -2.56&  -2.76*&   -.90\\
   89&  221&   7.79&  -1.20&  -1.59&  -1.10&   89&  222&   7.14&   1.37&   1.22&    .60\\
   89&  223&   6.79&   2.35&   1.99&   2.40&   89&  224&   6.32&   4.65&   4.55*&   5.60\\
   89&  226&   5.50&   8.64&   8.62&   9.20&   89&  227&   5.05&  10.89&  10.62&  11.00\\
   90&  217&   9.43&  -5.12&  -5.59*&  -3.60&   90&  219&   9.52&  -5.33&  -5.84&  -6.00\\
   90&  220&   8.96&  -4.99&  -4.79&  -5.00&   90&  221&   8.64&  -3.06&  -3.58*&  -2.50\\
   90&  222&   8.13&  -2.65&  -2.47&  -2.50&   90&  223&   7.58&    .18&   -.32&    .00\\
   90&  224&   7.31&    .05&    .22&    .00&   90&  225&   6.92&   2.57&   2.07&   3.00\\
   90&  229&   5.17&  10.98&  10.61&  11.60&91&  215&   8.17&  -1.62&  -1.82&  -1.80\\
   91&  217&   8.50&  -2.60&  -2.85&  -2.30&   91&  218&   9.80&  -5.62&  -5.75*&  -3.70\\
   91&  222&   8.70&  -2.82&  -2.98*&  -1.90&   91&  223&   8.35&  -2.16&  -2.51&  -1.80\\
   91&  224&   7.63&    .46&    .34&    .00&   91&  225&   7.40&    .92&    .57&    .40\\
   91&  226&   6.99&   2.77&   2.67&   2.40&   91&  227&   6.59&   4.05&   3.72&   3.70\\
   91&  229&   5.84&   7.51&   7.21&   8.10&   92&  226&   7.57&   -.05&    .16&   -.30\\
   92&  228&   6.81&   2.80&   3.01&   2.90&   92&  229&   6.48&   5.26&   4.82&   4.30\\
   92&  231&   5.56&   9.77&   9.40&   9.80&   93&  229&   7.01&   3.21&   2.91&   2.30\\
   93&  231&   6.37&   5.91&   5.62&   5.40&   93&  235&   5.20&  12.09&  11.86&  12.60\\
   94&  232&   6.72&   4.05&   4.29&   4.00&   94&  233&   6.42&   6.43&   6.04&   6.00\\
   94&  234&   6.32&   5.83&   6.06&   5.80&   94&  235&   5.96&   8.64&   8.26&   7.70\\
   94&  237&   5.75&   9.73&   9.34*&  11.10&   94&  239&   5.24&  12.66&  12.30&  12.00\\
   95&  238&   6.05&   8.71&   8.75&   9.30&   95&  239&   5.92&   9.03&   8.76&   8.70\\
   95&  240&   5.71&  10.49&  10.53&  11.00&   95&  242&   5.60&  11.10&  11.13&  11.70\\
   95&  243&   5.44&  11.68&  11.37&  11.40&   96&  238&   6.62&   5.37&   5.61&   4.90\\
   96&  240&   6.40&   6.37&   6.59&   6.50&   96&  241&   6.19&   8.44&   8.04&   8.60\\
   96&  243&   6.18&   8.49&   8.05&   9.00&   96&  244&   5.91&   8.79&   8.98&   8.80\\
   97&  245&   6.46&   7.31&   6.99*&   9.20&   97&  247&   5.89&  10.15&   9.84*&  10.90\\
   98&  242&   7.52&   2.55&   2.78&   2.40&   98&  244&   7.34&   3.24&   3.43&   3.10\\
   98&  250&   6.14&   8.55&   8.73&   8.60&   98&  251&   6.18&   9.42&   8.96*&  10.90\\
   98&  252&   6.22&   8.15&   8.27&   8.00&   98&  253&   6.12&   9.72&   9.23&   8.70\\
   98&  254&   5.94&   9.59&   9.71&   9.30&   99&  251&   6.61&   7.52&   7.19&   7.40\\
   99&  252&   6.76&   7.18&   7.11&   7.60&   99&  253&   6.75&   6.88&   6.50&   6.20\\
   99&  255&   6.44&   8.31&   7.93&   7.60&  100&  251&   7.43&   4.77&   4.32*&   6.00\\
  100&  252&   7.15&   4.83&   5.00&   4.90&  100&  253&   7.20&   5.69&   5.21*&   6.70\\
  100&  254&   7.31&   4.18&   4.31&   4.10&  100&  255&   7.24&   5.52&   5.02&   4.80\\
  100&  256&   7.03&   5.33&   5.44&   5.10&  100&  257&   6.87&   7.08&   6.56&   6.80\\
  102&  254&   8.23&   1.58&   1.78&   1.70&  102&  255&   8.45&   1.91&   1.44*&   2.80\\
  102&  256&   8.59&    .39&    .55&    .50&  102&  257&   8.46&   1.87&   1.37&   2.10\\
  102&  259&   7.81&   4.13&   3.63&   4.00&  106&  263&   9.40&    .39&   -.02&    .00\\ \hline

                 &VSS&DF&DF&DF\\
                &       &C=1.0&C=0.9&C=0.8\\ \hline
$\chi^2/F$&8.2&7.8&6.0&26.9 \\ 
\end{tabular} 
* indicates theoretical logarithmic half lives which differ by more than one unit from the experimental data.
\end{table}

The estimates of the logarithmic $\alpha$-decay half lives using the Viola-Seaborg relationship with Sobiczewski constants (eqn.(24), eqn.(25) and eqn.(26) of reference \cite{r19}) presented in the table-5 above for comparison with present calculations with DF potentials also show that the results of present calculations are slightly better. The degree of reliability of the present SAFM estimates using DF potentials for the $\alpha$ decay lifetimes are also better than the liquid drop description \cite{r16,r17} for most of the cases. In the present illustrative calculations the complete set of experimental data of references \cite{r4,r16,r18} for the $\alpha$ decay half lives have been chosen for comparison with the present theoretical calculation for which the experimental ground state masses  \cite{r20} for the parent and daughter nuclei are available. This ensures that there is then no uncertainty in the determination of the released energy which is one of the crucial quantities for the quantitative estimates for the decay half lives. Those results of logarithmic half lives which differ by more than one unit from the experimental ones are all odd-A or odd-odd $\alpha$-emitters. All odd-odd or odd-A nuclei have non-zero spins and atleast for some cases $\alpha$-particles have to  carry a minimum angular momentum required to satisfy the spin-parity conservation and thus requiring a minimum centrifugal barrier which will enhance the theoretical half-lives. Also for these cases there are considerable effects of deformation on the potential. 

      The half lives for cluster radioactivity have been analyzed with various microscopic and phenomenological nuclear potentials within the SAFM. The microscopic DF potentials are based on profound theoretical basis. The first turning point, which depends critically on nuclear potential, are quite different for different nuclear potentials while the second turning points are same as it is due to Coulomb potential. The results of the present calculations with microscopic DF potential within the SAFM are in good agreement to the experimental data. Although the first turning points and the potentials between the turning points are very different for the parabolic potential of  ASAFM and the microscopic DF potential, the results for the half lives are not much different. Half life estimates with nuclear proximity potential or other phenomenological potentials used within the SAFM do not agree with experimental results. The parabolic shape of the barrier adopted by the ASAFM \cite{r4} is difficult to justify in physical terms. The present study replaces the phenomenological parabolic barrier shape of ASAFM by a microscopic calculation of the interaction potential and puts the empirical approach of SAFM on firm theoretical basis.

\end{document}